\documentclass[superscriptaddress,
amsmath,amssymb,aps,prc,
showkeys,
twocolumn,nofootinbib
]{revtex4-1}
\usepackage{amssymb}

\usepackage{amsmath,mathrsfs,amssymb}
\usepackage{eurosym}
\usepackage{graphicx,subfigure}
\usepackage{color}
\usepackage{indentfirst}
\usepackage{extarrows}
\usepackage{hyperref}
\hypersetup{colorlinks=true, citecolor=blue, urlcolor=blue, linkcolor=blue}
\usepackage{ulem}
\usepackage{bm}
\usepackage[latin1]{inputenc}
\usepackage[title]{appendix}
\usepackage[T1]{fontenc}

\begin{document}

\title{ \bf{ Proton $0.01$ MeV resonance width  and low-energy $S$-factor of $p+ {}^{10}{\rm  B}$ fusion}}
\author{A. M. Mukhamedzhanov}
\email{akram@comp.tamu.edu}
\affiliation{Cyclotron Institute, Texas A$\&$M University, College Station, TX 77843, USA}

\begin{abstract}
{\bf{Background}:} The  ${}^{10}{\rm B}(p,\alpha){}^{7}{\rm Be}$ reaction is of interest for nuclear reaction theory, nuclear astrophysics and is important for neutronless (aneutronic) fusion  ${}^{11}{\rm B}(p,\,2\,\alpha){}^{4}{\rm He}$. At low-energies  the  $\,S$-factor  of the  ${}^{10}{\rm B}(p,\alpha){}^{7}{\rm Be}$ reaction  is contributed  by the near-threshold resonance $(E_{x}=8.70\,{\rm MeV};\,\frac{5}{2}^{+})$ with the resonance energy $E=0.01\,$MeV  and by higher resonances. Contrary to the $\alpha$-width, the proton resonance width of this resonance is unknown.\\
{\bf{Purpose}:} In this paper, the proton resonance width of the near-threshold resonance is calculated using two different approaches. The values of the proton width are used to calculate the low-energy $S$-factor. \\
{\bf{Method:}} First, the proton resonance width is estimated using the mirror symmetry of the resonance  $ {}^{11}{\rm C}(E_{x}=8.70\, {\rm MeV};\,\frac{5}{2}^{+})$ and the mirror bound state  ${}^{11}{\rm B}(E_{x}=9.272\, {\rm MeV};\,\frac{5}{2}^{+})$. In the second approach, this width is estimated using the $R$-matrix definition of the observable resonance width, which is expressed in terms of the $p-{}^{10}{\rm B}$ resonant wave function calculated in the potential approach and the spectroscopic  factor. \\
{\bf{Results}:} Depending on the method chosen, the calculated proton resonance width varies from  $\,1.03 \times 10^{-19}\,$MeV to $\,2.96 \times 10^{-19}\,$MeV. The role of the near-threshold resonance is determined using fitting of two low-energy $S$-factors from direct measurements [C. Angulo {\it et al.}, Z. Phys. A {\bf 345}, 333 (1993)]  and from the indirect Trojan horse method (THM) [ A. Cvetinovi\'c {\it et al.}, Phys. Rev. C {\bf 97}, 065801 (2018)].
Within the framework of the $R$-matrix method  using the determined proton resonance widths and the modified THM parameters for six low-lying resonances, the low-energy  $S$-factors were calculated and compared with the corresponding experimental  $S$-factors. The closest agreement with the data is achieved with the proton resonance widths $1.0 \times 10^{-19}$  MeV when fitting the $S$-factor from the THM indirect measurements, and  $\,1.68 \times 10^{-19}\,$ MeV and  $\,2.5 \times 10^{-19}\,$ MeV when fitting the $S$-factor  
from  [Angulo {\it et al}, Z. Phys. A {\bf 345}, 333 (1993)].\\
{\bf{Conclusion}:}  
Our calculated proton resonance widths using the mirror symmetry and  the $R$-matrix  method  are an order of magnitude larger than the 
phenomenological $S$-factor determined  from the $R$-matrix fit of the latest measurements [ Van de Kolk  {\it et al.}, Phys. Rev. C {\bf 105}, 055802 (2022)]. Using the theoretically determined  proton resonance widths we achieved excellent fits of the low-energy $S$-factors determined from the direct and indirect measurements.
   
\end{abstract}

\maketitle


\section{\bf{Introduction}}

\begin{figure}[tbp]
\includegraphics*[width=\linewidth]{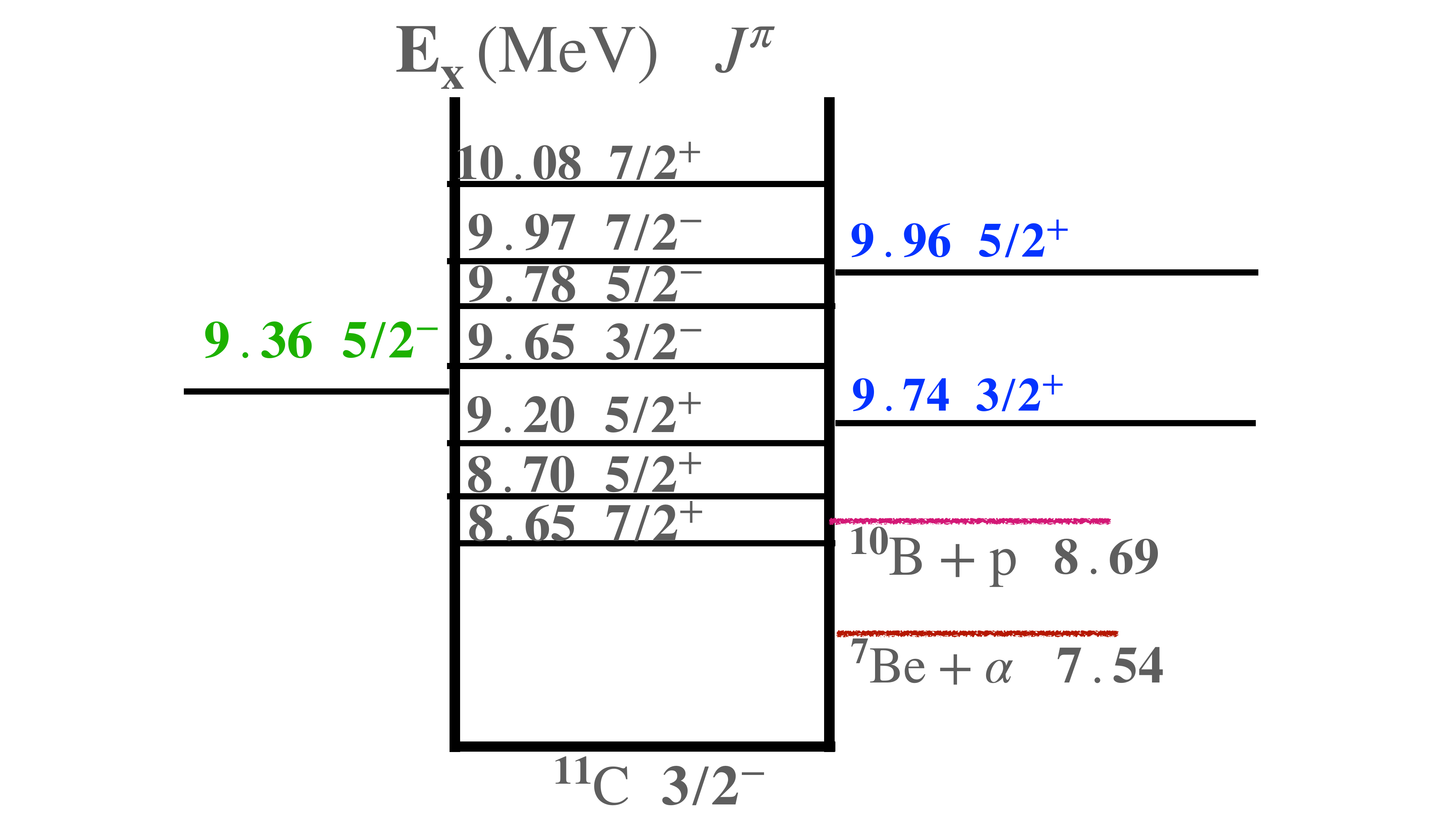}
\caption{The energy levels of ${}^{11}{\rm C}$ contributing to low-energy    ${}^{10}{\rm B}(p,\alpha){}^{7}{\rm Be}$ 
reaction cross-section. $E_{x}$ is the excitation energy expressed in MeV. Only levels slightly above the proton threshold and one subthreshold bound state in the channel ${}^{10}{\rm B}+p$  are shown. The black numbers are data taken from \cite{Kelley}. The  blue data are from the recent $R$-matrix fit to the $S$-factor and cross-section of the   ${}^{10}{\rm B}(p,\alpha){}^{7}{\rm Be}$   reaction  reported in \cite{Kolk}.  The green numbers are taken from \cite{Spitaleri2018}.
The red lines are the proton and $\alpha$-channel thresholds. }
\label{fig_energylevels}
\end{figure}

The  ${}^{10}{\rm B}(p,\alpha){}^{7}{\rm Be}$  reaction at low energy is of interest to nuclear physics, nuclear astrophysics, and applied physics. This reaction allows nuclear physicists to investigate low-energy  ${}^{11}{\rm C}$ resonances at excitation energies $E_{x} \geq 8.70$ MeV,  which are above the proton and $\alpha$  decay channels.  There are controversies about the location and spin-parity assignments of low-energy ${}^{11}{\rm C}$ resonances. Fig. \ref{fig_energylevels}  depicts the low-energy resonances taken from compilation \cite{Kelley}, from the latest $R$-matrix fit to the experimental cross-section of the  ${}^{10}{\rm B}(p,\alpha){}^{7}{\rm Be}$  reaction and from the indirect THM \cite{Spitaleri2018}.
One can see that the $R$-matrix fit to direct measurements \cite{Kolk}  in the energy interval  $E= 0.73-1.82$ MeV, where $E \equiv E_{p{}^{10}{\rm B}}$ is the $p-{}^{10}{\rm B}$ relative kinetic energy,  required change of two resonance assignments given in compilation \cite{Kelley}. In particular, the level $\frac{3}{2}^{+}$ just above the level $(8.70\,{\rm MeV};\,\frac{5}{2}^{+})$  is quite plausible because a similar level is observed in the mirror nucleus ${}^{11}{\rm B}$.
However, in the recently published paper \cite{Okolowicz}  within the framework of the shell model embedded in the continuum (SMEC) was confirmed the presence of the level $(E_{x}= 9.20\,{\rm MeV}; \frac{5}{2}^{+})$ rather than the level  
$(E_{x}= 9.74\,{\rm MeV}; \frac{3}{2}^{+})$. Besides, the THM experiment \cite{Spitaleri2018} suggested that there is a new level, 
$(E_{x}= 9.36\,{\rm MeV}; \frac{5}{2}^{-})$, which was not mentioned in the literature before.  

This uncertainty regarding the low-level resonances in ${}^{11}{\rm C}$ requires  further analysis  because  they affect the low-energy cross-section and  $S$-factor of the  ${}^{10}{\rm B}(p,\alpha){}^{7}{\rm Be}$  reaction.  It is the reason why in this paper, we constrained our fit by  $E <0.3$ MeV, where the near-threshold resonance $(E_{x}= 8.70\,{\rm MeV}; \frac{5}{2}^{+})$ dominates.

In nuclear astrophysics, the  ${}^{10}{\rm B}(p,\alpha){}^{7}{\rm Be}$ reaction represents the main ${}^{10}{\rm B}$ destruction channel in H-rich main-sequence outer layers of stars with the Gamow window energy $0.01 \pm 0.005$ MeV for such stellar environments [3]. The  $\,\frac{5}{2}^{+}\,$ resonance at $E_{R}=0.01\, {\rm MeV}\,$  corresponding to the excitation energy  $E_{x}=8.70\, {\rm MeV}\,$  of ${}^{11}{\rm C}$ (see Fig. \ref{fig_energylevels}) dominates the ultra low-energy cross-section of the  ${}^{10}{\rm B}(p,\alpha){}^{7}{\rm Be}$ reaction. 
This resonance is key in predicting boron abundances and constraining mixing phenomena in such stars \cite{Boesgaard}.

Another role of the ${}^{10}{\rm B}(p,\alpha){}^{7}{\rm Be}$  reaction is related to neutronless  (also called aneutronic synthesis \cite{Kolk}) laser-driven, hot-plasma fusion, which attracts increasing attention as a new energy source that would be free of radioactive products with large half-life. One of  the sources  of the  aneutronic  fusion is the   ${}^{11}{\rm B}(p,\,2\,\alpha){}^{4}{\rm He}$ $\,(Q=8.7\,{\rm MeV})$ reaction.  Natural boron  contains $\sim 80\%$  of the ${}^{11}{\rm B}$ isotope  and  $\sim 20\%$ of ${}^{10}{\rm  B}$.  The presence of the isotope ${}^{10}{\rm B}$  results in the collateral reaction  ${}^{10}{\rm B}(p,\alpha){}^{7}{\rm Be}\,$ ($Q=1.152\,$MeV), which can contaminate the fusion reactors due to the generation of the radioactive ${}^{7}{\rm Be}$.  
But the ${}^{10}{\rm B}(p,\alpha){}^{7}{\rm Be}$  reaction plays a positive role in providing an additional tool to measure the temperature of hot plasma used in ignition facilities (see \cite{Kolk} and references therein). Note that very low energies ($\sim 0.01 \,{\rm MeV}$)  is the operational regime of the  National Ignition Facility (NIF). 

The most advanced measurements of the ${}^{10}{\rm B}(p,\alpha){}^{7}{\rm Be}$ $S$-factor  were reported  in  Refs.\cite{Angulo,Spitaleri2014,Spitaleri2017} and \cite{Kolk}.  The work \cite{Angulo}  presented the lowest direct measurements of the $S$-factor of the  ${}^{10}{\rm B}(p,\alpha){}^{7}{\rm Be}$  reaction.  Other three papers, \cite{Spitaleri2018,Spitaleri2014,Spitaleri2017}, addressed the indirect measurements of the $S$-factor in the energy region $E=0.005 - 1.5$ MeV using the THM. It is the lowest measured energy interval; however, the uncertainty of the measured $S$-factor quickly increases as the energy decreases. The most accurate  direct measurements  of the ${}^{10}{\rm B}(p,\alpha){}^{7}{\rm Be}$  cross-section and $S$-factor were recently reported in \cite{Kolk} in which the ${}^{10}{\rm B}(p,\alpha){}^{7}{\rm Be}$  cross-section was measured in the proton energy interval $0.8-2.0$ MeV.  It is an excellent work in which the uncertainties of the measured cross-section and $S$-factors were reduced to $\approx 10\%$ in the measured energy interval.  

The very low-energy cross-section of  ${}^{10}{\rm B}(p,\alpha){}^{7}{\rm Be}$, contributed by the $E_{R}=0.01$  MeV resonance,  depends on the proton resonance width in the entry channel and the $\alpha$-particle resonance width in the exit channel of the reaction. While the latter is well established,  $\Gamma_{\alpha}= 0.015$ keV   \cite{Kelley},  the proton resonance width, which could be crucial for the   ${}^{10}{\rm B}(p,\alpha){}^{7}{\rm Be}$  cross-section and $S$-factor evaluations, is the subject to large uncertainty.

In this paper, I will present a theoretical evaluation of the proton resonance width of the $E_{R}=0.01$ MeV resonance 
using two different theoretical approaches,  mirror symmetry and $R$-matrix, and check an impact of the proton resonance width on the low-energy $S$-factor of the ${}^{10}{\rm B}(p,\alpha){}^{7}{\rm Be}$ reaction.
Both methods give results by about an order of magnitude larger than the phenomenological proton resonance width obtained in \cite{Kolk} using the $R$-matrix fitting to the higher-energy experimental data. Obtained higher proton resonance widths allowed us to fit the low-energy $S$-factors from  \cite{Angulo} and \cite{Spitaleri2018}. The paper uses the system of units in which $\,\hbar=c=1\,$.\\
 \newline
  
\section{\bf{ Proton resonance width from mirror symmetry  of   $\mathbf{{}^{11}{\rm {\bf C}}(E_{x}=8.70\, {\rm {\bf MeV}})}$ and  $\mathbf{{}^{11}{\rm {\bf B}}(E_{x}=9.272\,  {\rm {\bf MeV}})}$}}
\label{MirrSym1}

The width of a narrow resonance can be expressed in terms of the ANC of the Gamow-Siegert wave function \cite{mukRes2023}. That is why we can extend the relationship between the ANCs of mirror bound states to the relationship between resonance widths and ANCs of the mirror nuclei \cite{timofeyuk2003,muk2019}. The calculated resonance widths and the ANCs depend strongly on the choice of the nucleon-nucleon (NN) force, but the ratios of the resonance widths and the ANCs for mirror bound states should not depend on the adopted NN force, see for details review \cite{mukblokh2022}.  

Another important feature of the mirror nuclei is the similarity of the internal mirror wave functions. Let us consider a mirror pair in a two-body potential model: $\,B_{1} =(a_{1}\,A_{1})\,$ in the resonance state and the loosely bound nucleus $\,B_{2}=(a_{2}\,A_{2})$. The mirror resonance state is obtained by replacing one of the neutrons with a proton. The additional Coulomb interaction pushes the bound-state level into a resonance level. 
The resonance and binding energy of the mirror states are significantly smaller than the depth of the nuclear potential. The Coulomb interaction is almost a constant in the nuclear interior. Hence, in the nuclear interior, which is all that matters to determine the ratio of the resonance width and the ANC of the mirror state,  the radial behavior of the mirror wave functions, which are both real, is very similar, and they differ only by normalization. In the outer region, the resonant and bound-state wave functions vary significantly.

In the case under consideration,  we consider the resonance state   ${}^{11}{\rm C}=(p\,{}^{10}{\rm B})[E_{x}=8.70\, {\rm MeV};\,\frac{5}{2}^{+}]\,$ and the mirror bound state  ${}^{11}{\rm B}=(n\,{}^{10}{\rm B})[E_{x}=9.272\, {\rm MeV};\,\frac{5}{2}^{+}]$. 
Following  \cite{muk2019},  we  express the ratio of the resonance  width $\Gamma_{ (p)\,l_{p}\,j_{p} }$  of the resonance state of ${}^{11}{\rm C}$  and the ANC $C_{(n)\,l_{n}\,j_{n}}$ of the mirror bound state  ${}^{11}{\rm B}$ \cite{mukRes2023} in terms of the ratio of the Wronskians ${\cal W}$:
 \begin{widetext}
\begin{align}
& \frac{ \Gamma_{(p)\,l_{p}\,j_{p} }}{ (C_{(n)\,l_{n}\,j_{n}})^{2}}\, =
\sqrt{\frac{2\,E}{\mu_{pA}}}\, \frac{\Big|\kappa_{n\,A}\,{\cal W}[I_{{l_{p}}\,{j_{p}}}(k_{p\,A},\,{r_{p\,A } }),\,F_{l_{p}}(k_{p\,A},\,r_{p\,A})]\Big|^{2}\,\Bigg|_{r_{p\,A}=R_{ch}}}{\Big|k_{p\,A}\,{\cal W}[I_{{l_{n}}\,{j_{n}}}(\kappa_{n\,A},\,r_{n\,A} ),\,F_{l_{n}}(i\,\kappa_{n\,A},\,r_{n\,A})]\Big|^{2}\,\Bigg|_{r_{n\,A}=R_{ch}}}                     
\label{GpANCratioOverlfunct1} \\
& \approx \frac{ {\mathtt S}_{ l_{p}\,j_{p} }} {{\mathtt S}_{l_{n}\,j_{n}  } }   \sqrt{\frac{2\,E}{\mu_{pA}}}\, \frac{\Big|\kappa_{n\,A}\,{\cal W}[\varphi_{{l_{p}}\,{j_{p}}}(k_{p\,A},\,{r_{p\,A } }),\,F_{l_{p}}(k_{p\,A},\,r_{p\,A})]\Big|^{2}\,\Bigg|_{r_{p\,A}=R_{ch}}}{\Big|k_{p\,A}\,{\cal W}[\varphi_{{l_{n}}\,{j_{n}}}(\kappa_{n\,A},\,r_{n\,A} ),\,F_{l_{n}}(i\,\kappa_{n\,A},\,r_{n\,A})]\Big|^{2}\,\Bigg|_{r_{n\,A}=R_{ch}}}.
\label{GammaANCmirrorratio2}
\end{align}
\end{widetext}
where the ratio $ \frac{ \Gamma_{(p)\, l_{p}\,j_{p}}}{ (C_{(n)\,l_{n}\,i_{nA}})^{2}}$ is dimensionless, $R_{ch}$ is the channel radius dividing the configuration space into the internal and external  regions. 
$I_{l_{p}\,j_{p}}$  is the  proton overlap function determining the projection of the bound-state wave function of nucleus $\,{}^{11}{\rm C}\,$ on  the channel  $p +{}^{10}{\rm B},\,$  and
$\,I_{l_{n}\, j_{n}}\,$ is the neutron overlap function determining the projection of the bound-state wave function of nucleus  $\,{}^{11}{\rm B}\,$ on the channel  $n+{}^{10}{\rm B}$.
$l_{p}$ and  $j_{p}$   ($l_{n}$ and $j_{n}$)  are the proton (neutron) orbital angular momentum and  total angular momentum in nucleus ${}^{11}{\rm C}$  (${}^{11}{\rm B}$).  For the case under consideration  $\, l_{p}=l_{n}=0,\,j_{p}=j_{n}=1/2.\,$
$E \equiv E_{pA}$  and $\mu_{pA}$ are the real part of the  resonance energy  and the reduced mass of the  $p-A$ pair, which
are expressed in MeV. $k_{pA}$ is the $p-A$ relative momentum and $\kappa_{n\,A}$ is the bound-state wave number of the bound state $(n\,A),\,$ $\,F_{l_{p}}(k_{p\,A},\,r_{p\,A})\,$  is the regular Coulomb solution of the $p+A$ scattering. 
$F_{l_{n}}(i\,\kappa_{n\,A},\,r_{n\,A})$ is the regular solution of the free  $n+A$ scattering.  

To obtain Eq. (\ref{GammaANCmirrorratio2})   we  use the single-particle approximation  in the internal  region 
\begin{align}
\,I_{{l_{N}}\,{j_{N}}}(r_{NA})= {\mathtt S}_{l_{N}\, j_{N} }^{1/2}\,\varphi_{{l_{N}}\,{j_{N}}}(r_{NA}).
\label{SFspappr1}
\end{align}
 Since the the Wronskian ratio (\ref{GpANCratioOverlfunct1})  is taken  at  the channel radius, we can use  the proton resonance wave 
 $\varphi_{{l_{p}}\,{j_{p}}}$   and the mirror neutron bound-state wave function   $\varphi_{{l_{n}}\,{j_{n}}}$, which are normalized to the unity over the internal region  ($r_{pA} \leq R_{ch}$  and $r_{nA} \leq R_{ch}$). 
$ {\mathtt S}_{l_{N}\, j_{N} }$  is the single-particle spectroscopic factor (SF)  of the $(N-A)_{l_{N}\, j_{N} }$.  

If the SFs of the mirror states are the same,  
\begin{align}
{\mathtt S}_{l_{p}\, j_{p} }= {\mathtt S}_{l_{n}\, j_{n} },
\label{SpSn1}
\end{align}
then   Eq. (\ref{GammaANCmirrorratio2})   reduces to
 \begin{widetext}
\begin{align}
& \frac{ \Gamma_{(p)\,l_{p}\,j_{p} }}{ (C_{(n)\,l_{n}\,j_{n}})^{2}}\, 
\approx   \sqrt{\frac{2\,E}{\mu_{pA}}}\, \frac{\Big|\kappa_{n\,A}\,{\cal W}[\varphi_{{l_{p}}\,{j_{p}}}(k_{p\,A},\,{r_{p\,A } }),\,F_{l_{p}}(k_{p\,A},\,r_{p\,A})]\Big|^{2}\,\Bigg|_{r_{p\,A}=R_{ch}}}{\Big|k_{p\,A}\,{\cal W}[\varphi_{{l_{n}}\,{j_{n}}}(\kappa_{n\,A},\,r_{n\,A} ),\,F_{l_{n}}(i\,\kappa_{n\,A},\,r_{n\,A})]\Big|^{2}\,\Bigg|_{r_{n\,A}=R_{ch}}}.
\label{GammaANCmirrorratio3}
\end{align}
\end{widetext}

Equations (\ref{GpANCratioOverlfunct1}), (\ref{GammaANCmirrorratio2})  and (\ref{GammaANCmirrorratio3}) allow one to determine the resonance width if the  ANC of the mirror bound state is known and vice versa.  The mirror wave functions,   $\,\varphi_{l_{p}\,j_{p}}(k_{pA},\,{r_{pA}})$  and    $\varphi_{{l_{n}}\,{j_{n}}}(\kappa_{nA},\,{r_{nA}})$,   are calculated using similar two-body $N-A$ nuclear potentials.   In the internal region ($r \leq  R_{ch}$), which is needed to calculate the right-hand-side of Eqs.  (\ref{GammaANCmirrorratio2})  and     (\ref{GammaANCmirrorratio3}),   both wave functions  are real and their radial behavior, due to the mirror symmetry, should look very similar.   
  
We can further simplify Eq. (\ref{GammaANCmirrorratio3}).
The Coulomb interaction varies very little in the nuclear interior, and its effect only leads to shifting the bound state's energy to the continuum. Hence, it can be assumed that $F_{l_{p}}(k_{pA},\,r_{pA})$  and $F_{l_{n}}(i\,\kappa_{nA},\,r_{nA})$ behave similarly in the nuclear interior except for the overall normalization, that is, 
\begin{align}
&F_{l_{p}}({k_{pA)}},\,{r_{pA}}) \approx \frac{ F_{l_{p}}(k_{pA},\,R_{ch})}{{{F_{{l_{n}}}}(i\,{\kappa _{nA}},\,{R_{ch}})}}\,F_{{l_{n}}}(i\,{\kappa _{nA}},\,{r_{nA}}).
\label{varphisimple1}
\end{align}

Neglecting further the difference between the mirror wave functions $\,\varphi_{l_{p}\,j_{p}}(k_{pA},\,{r_{pA}})$ and $\varphi_{{l_{n}}\,{j_{n}}}(\kappa_{nA},\,{r_{nA}})$  in the nuclear interior we obtain the approximate expression 
for $ \frac{ \Gamma_{(p)\, l_{p}\,j_{p} }}{ (C_{(n)\,l_{n}\,j_{n}})^{2}}\,$:
\begin{align}
& \frac{ \Gamma_{(p)\, l_{p}\,j_{p} }}{ (C_{(n)\,l_{n}\,j_{n}})^{2}} \approx\sqrt{\frac{2\,E}{\mu_{pA}}} \,{\Bigg| \frac{ F_{l_{p}}(k_{pA},\,R_{ch})}{{{F_{{l_{n}}}}(i\,{\kappa _{nA}},\,{R_{ch}})}}  
\Bigg|^2} .
\label{Ratioapprox2}
\end{align} 
This equation provides the easiest way to determine  $\,\frac{ \Gamma_{(p)\, l_{p}\,j_{p} }}{ (C_{(n)\,l_{n}\,j_{n}})^{2}}\,$ because to calculate it one needs to know only the Coulomb scattering wave function in continuum and free scattering wave function at imaginary momentum for  the mirror bound state. However,  Eq. (\ref{Ratioapprox2})  is less accurate than  Eq. (\ref{GammaANCmirrorratio2}).

\begin{figure}[tbp]
\includegraphics*[scale=0.5]{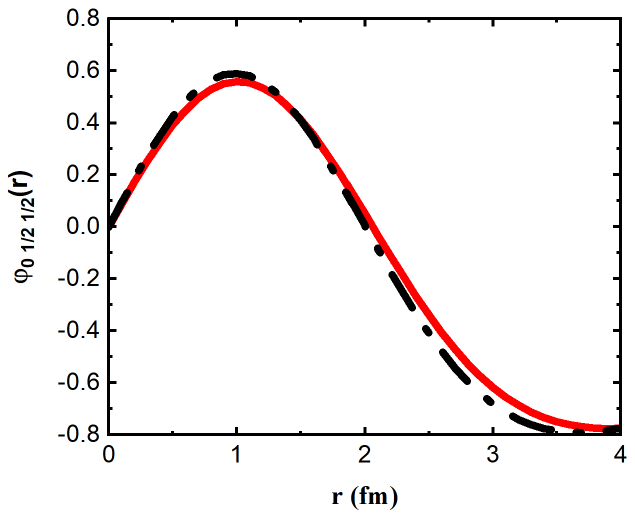}
\caption{Internal single-particle resonance (solid red line) and bound-state (black dash-dotted line)  wave functions of   ${}^{11}{\rm C}(E_{x}=8.70\, {\rm MeV};\,\frac{5}{2}^{+})$  and    ${}^{11}{\rm B}(E_{x}=9.272\, {\rm MeV};\,\frac{5}{2}^{+})$  states, respectively. Both wave functions are normalized to unity in the internal region $0 \leq  r \leq 4\,$fm. }
\label{fig_r04fm}
\end{figure}

\begin{figure}[tbp]
\includegraphics*[scale=0.5]{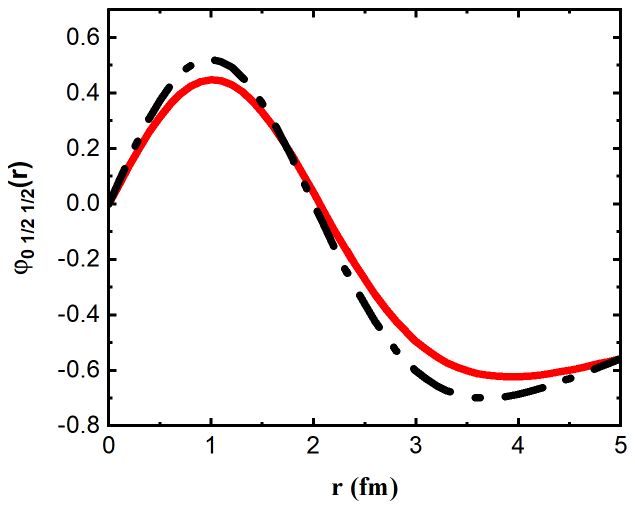}
\caption{ The same as in Fig. \ref{fig_r04fm} but for $0 \leq r \leq 5\,$fm.}
\label{fig_r05fm}
\end{figure}

Figs. \ref{fig_r04fm} and  \ref{fig_r05fm} show resonance and bound-state wave functions normalized in the internal regions $0 \leq r \leq 4\,$fm and   $0 \leq r \leq 5\,$fm, respectively.
The single-particle resonance and bound-state wave functions presented in these figures are calculated using the Woods-Saxon potential with the radial parameter $r_{0}=1.27\,$fm and diffuseness $a=0.67\,$fm \cite{Okolowicz}. 
The depth of the potential is $V_{0}=61.7065\,$MeV for the resonance state, and  $V_{0}=63.0952\,$MeV for the mirror bound state. Since the nucleon orbital angular momentum in the mirror states  ${}^{11}{\rm C}=(p\,{}^{10}{\rm B})[E_{x}=8.70\, {\rm MeV};\,\frac{5}{2}^{+}]\,$ and  ${}^{11}{\rm B}=(n\,{}^{10}{\rm B})[E_{x}=9.272\, {\rm MeV};\,\frac{5}{2}^{+}]$ is $l_{N}=0$, spin-orbital interaction vanishes. 
We inspect that the normalized in the internal regions mirror wave functions with $R_{ch}=4,\,5\,$fm are very close reflecting mirror symmetry property \footnote[2]{The resonance and bound-state wave functions are normalized to unity and compared only in the internal region, where the resonance wave function is a real function and can be compared with the bound-state one. 
The $R$-matrix method employs such normalized resonant wave functions to determine the resonance width.}.

\begin{figure}[tbp]
\includegraphics*[scale=0.5]{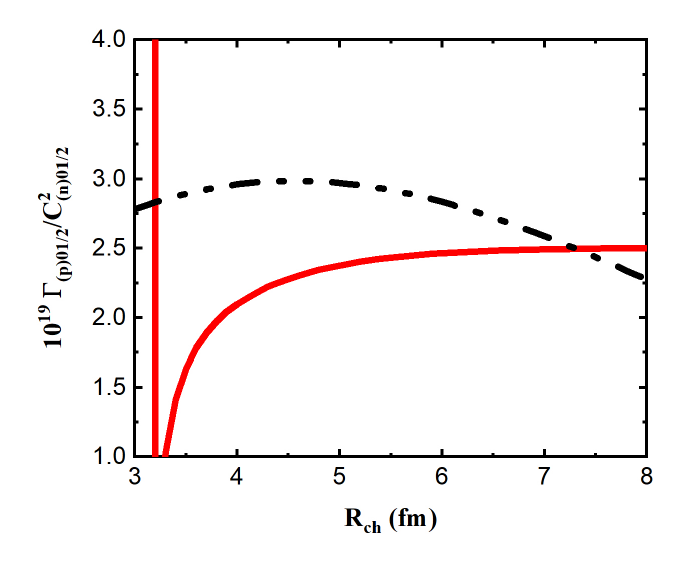}
\caption{ Ratio $\frac{ \Gamma_{(p)\, 0\,1/2}}{ (C_{(n)\,0\, \frac{1}{2} })^{2}}$ determined from mirror symmetry of ${}^{11}{\rm C}(E_{x}=8.70\,{\rm MeV};\,\frac{5}{2}^{+})\,$ and  ${}^{11}{\rm B}(E_{x}=9.27\,{\rm MeV};\,\frac{5}{2}^{+})\,$  states. The solid red line is determined using Eq. (\ref{GammaANCmirrorratio3}), and the black dash-dotted-dotted line is determined using Eq. (\ref{Ratioapprox2}). Note that a discontinuity of the solid red line is caused by the node of the bound-state wave function at  $r \approx 2$ fm, see Figs. \ref{fig_r04fm} and \ref{fig_r05fm}. }              
\label{fig_GpCnmirsym1}
\end{figure}

Fig.  \ref{fig_GpCnmirsym1}  shows the ratio  $\,\frac{ \Gamma_{(p)\,0\, \frac{1}{2} }}{ (C_{(n)\,0\, \frac{1}{2} })^{2}}$  determined from Eqs. (\ref{GammaANCmirrorratio3}) and (\ref{Ratioapprox2}).  
From  Eq. (\ref{GammaANCmirrorratio3}) we get  $\frac{ \Gamma_{(p)\,0\, \frac{1}{2} }}{ (C_{(n)\,0\, \frac{1}{2}})^{2}}= 1.5\times 10^{-19}$  MeVfm for $R_{ch}=4$ and $5$ fm.  Less accurate value is determined from Eq. (\ref{Ratioapprox2}):  
$\frac{ \Gamma_{(p)\,0\, \frac{1}{2}}}{ (C_{(n)\,0\, \frac{1}{2}})^{2}}= 2.4\times 10^{-19}$  MeVfm at $R_{ch}=4.6$ fm corresponding to the peak  
of the ratio $\frac{ \Gamma_{(p)\, 0\, \frac{1}{2} }}{ (C_{(n)\,0\, \frac{1}{2} })^{2}}$.  

To determine the proton resonance width, we need the neutron ANC  for the virtual breakup ${}^{11}{\rm B}[E_{x} =9.27\,{\rm MeV};\,\frac{5}{2}^{+}]  \to n + {}^{10}{\rm B}.\,$ 

\subsubsection{ \bf{ Shell-model ANC of neutron removal from}  $\mathbf{  {}^{11}{\rm B}[E_{x} =9.27\,{\rm MeV};\,\frac{5}{2}^{+}] }$  }
\label{ANCSHM1}

One can get this  ANC by multiplying the single-particle ANC by the square root of the neutron's SF. 
However, the SF is not an observable quantity \cite{frunsthamm2002,mukkad2010} and depends on the method and model used to determine it.

First, we use the SF determined  from the shell model and  the single-particle potential  from \cite{Okolowicz}. 
The single-particle neutron ANC is determined as the amplitude of the tail of the neutron bound-state wave function normalized to unity in the entire coordinate space  \cite{mukblokh2022}. This wave function is calculated in the above-presented Woods-Saxon potential. The estimated in \cite{Okolowicz} the proton's SF  of the resonance state ${}^{11}{\rm C}=(p\,{}^{10}{\rm B})[E_{x}= 8.70\,{\rm MeV};\,\frac{5}{2}^{+}]\,$ is $0.33$, which is close to the phenomenological SF ${\mathtt S}_{p}=0.31\,$ determined in \cite{Kolk} from the $R$-matrix fitting to the low-energy experimental data.  Owing to the mirror symmetry of the resonance state $(p\,{}^{10}{\rm B})[E_{x}= 8.70\,{\rm MeV};\,\frac{5}{2}^{+}]$ and the bound state $(n\,{}^{10}{\rm B})[E_{x}= 9.27\,{\rm MeV};\,\frac{5}{2}^{+}]$  I assume the same SF for both states. 
Then the ANC  for the virtual breakup ${}^{11}{\rm B}[E_{x} =9.27\,{\rm MeV};\,\frac{5}{2}^{+}] \to n + {}^{10}{\rm B}\,$ is $C_{n}= 1.114$ fm${}^{-1/2}$. Using this ANC, we find from Eqs. (\ref{GammaANCmirrorratio3})  and  (\ref{Ratioapprox2})  
$\,\Gamma_{(p)\,0\, \frac{1}{2}}= 1.87\times 10^{-19}$ MeV for $R_{ch}=4.0$ and $5.0$ fm  and less accurate $\,\Gamma_{(p)\,0\, \frac{1}{2}}= 2.96\times 10^{-19}\,$ MeV for $R_{ch}= 4.6\,$ fm,  respectively. 

\subsubsection{\bf{ ANC of } of neutron removal from $\mathbf{{}^{11}{\rm B}[E_{x} =9.27\,{\rm MeV};\,\frac{5}{2}^{+}]}\,$  \bf {from  stripping  reactions}   } 
\label{ANCStr1}
 
In Refs. \cite{HindsMidleton}  and \cite{Fortune} the  nucleon stripping reactions ${}^{10}{\rm B}({}^{3}{\rm He},\,d){}^{11}{\rm C}$  and  ${}^{10}{\rm B}(d,p){}^{11}{\rm B}$  were studied at low energies.  The differential cross sections  of  the
reactions populating   $\,8.65-8.69\,$ MeV  and $\,9.19-9.27\,$  MeV  doublets  in ${}^{11}{\rm C}$ and ${}^{11}{\rm B}$ were measured.  The determined SFs  of the mirror  $8.69$  \footnote[2]{Contemporary value of the excitation energy of this level  is  $8.70$ MeV.}  MeV in   ${}^{11}{\rm C}$   and $9.27$  MeV  in ${}^{11}{\rm B}$  are  $0.2$  and   $0.17$, respectively.
For the neutron bound-state potential parameters in ${}^{11}{\rm B}$  the single-particle ANC  is  $2.0$ fm${}^{-1/2}$. 
Then the ANC from the ${}^{10}{\rm B}(d,p){}^{11}{\rm B}$  is  $0.83$ fm${}^{-1/2}$.  Since the proton SF 
of the  ${}^{11}{\rm C}$   is slightly higher than the neutron one to determine the resonance width, we are supposed to use Eq. (\ref{GammaANCmirrorratio2}).  However,  there is no explanation in Ref. \cite{HindsMidleton}   how transfer to resonance was 
calculated and how the resonance wave function was normalized.  That is why I still assume the SFs for the mirror states under consideration are the same.  

Taking into account that from  Eq. (\ref{GammaANCmirrorratio3}) we get  $\frac{ \Gamma_{(p)\,0\, \frac{1}{2} }}{ (C_{(n)\,0\, \frac{1}{2}})^{2}}= 1.5\times 10^{-19}$  MeVfm for $R_{ch}=4$ and $5$ fm and that the neutron ANC from the ${}^{10}{\rm B}(d,p){}^{11}{\rm B}$ 
reaction is $0.83$ fm${}^{-1/2}$  we get  $\,\Gamma_{(p)\,0\, \frac{1}{2}}= 1.03\times 10^{-19}$ MeV for $R_{ch}=4$ and $5$ fm.

 Less accurate value  of the ratio $\frac{ \Gamma_{(p)\,0\, \frac{1}{2}}}{ (C_{(n)\,0\,1/2})^{2}}= 2.4 \times 10^{-19}$  MeVfm at $R_{ch}=4.6$ fm  obtained from Eq. (\ref{Ratioapprox2})  and the proton  ANC  $0.83$ fm${}^{-1/2}$ lead to  $\,\Gamma_{(p)\,0\, \frac{1}{2}}= 1.65 \times 10^{-19}$ MeV.
       
\section{\bf{Proton resonance width of resonance at $\mathbf{E_{R}=0.01}$ MeV  from $\mathbf{R}$-matrix  approach}}
\label{Rmatrix}

In the $R$-matrix method the observable proton resonance width is determined by
\begin{align}
 \Gamma_{(p)\,0\,  \frac{1}{2} }= 2\,P_{0}(E_{R}, R_{ch})\,\gamma_{(p)\,0\,  \frac{1}{2}  }^{2}.
 \label{GpRmtrx1}
 \end{align}
 Here  $P_{0}(E_{R}, R_{ch})$ is the penetrability factor in the $l_{p}=0$ partial wave,
 \begin{align}
 \gamma_{(p)\,0\,  \frac{1}{2} }^{2}= {\mathtt S}_{(p)\,0\,  \frac{1}{2} }\,\frac{ {\tilde  \gamma}_{(p)\,0\,  \frac{1}{2}  }^{2} }{ 1+ {\tilde  \gamma}_{(p)\,0\,  \frac{1}{2}  }^{2}\,\frac{ {\rm d}{\hat S}_{0}(E) }{ {\rm d}E}\Big|_{E=E_{R}}  }
\label{gammaobs1}
\end{align}
is the observable reduced width of the resonance state (see Appendix),  ${\hat S}_{0}(E)$ is the $R$-matrix shift factor in the $l_{p}=0$ partial wave.
\begin{align}
 {\tilde  \gamma}_{(p)\,0\, \frac{1}{2} }^{2} = \frac{\varphi_{0\, \frac{1}{2}}^{2}(R_{ch})}{2\,\mu_{pA}\,R_{ch}}
\label{formgamma1}
\end{align}
is the formal $R$-matrix reduced width, $\varphi_{0\,\frac{1}{2}}(r_{pA})$ is the  internal  resonance radial  wave function normalized 
to unity in the internal region  $\,0 \leq r \leq  R_{ch}.\,$
 For $R_{ch}=4.0$ fm we get $\varphi_{0\,\frac{1}{2}}(4.0\, {\rm fm})= -0.775$ fm${}^{-1/2},\,$   $\, {\tilde  \gamma}_{0\,\frac{1}{2}}= -1.851$ MeV${}^{1/2}.\,$  For the proton SF ${\mathtt S}=0.31\,$  we get 
 $ \gamma_{(p)\,0\,\frac{1}{2}} = - 0.64$ MeV${}^{1/2}$. The penetrability factor $\,P_{0}(0.01\,{\rm MeV},\, 4.0\,{\rm fm} )= 2.05\times 10^{-19}$ and  $ \Gamma_{(p)\,0\,\frac{1}{2}}= 1.683\times 10^{-19}$ MeV. 

For $R_{ch}=5.0$ fm we get $\varphi_{0\,\frac{1}{2}}(5.0\,{\rm fm})= - 0.557$ fm${}^{-1/2},\,$   $\, {\tilde  \gamma}_{0\,\frac{1}{2}}= -1.190\,$ MeV${}^{1/2}$ 
 and  $ \gamma_{(p)\,0\,\frac{1}{2} } = - 0.483$ MeV${}^{1/2}$. The penetrability factor $P_{0}(0.01\,{\rm MeV},\, 5.0\,{\rm fm} )= 3.99\times 10^{-19}$ and  $ \Gamma_{(p)\,0\,\frac{1}{2}}= 1.863\times 10^{-19}$ MeV. 
 
 \section{\bf{Discussion of the obtained proton resonance width}}
 
 The summary of the obtained in this paper theoretical proton resonance widths
 of the near-threshold resonance ${}^{11}{\rm C}=(p\,{}^{10}{\rm B})[E_{x}=8.70\, {\rm MeV};\,\frac{5}{2}^{+}]\,$  using the mirror symmetry and the $R$-matrix approach are presented in Table \ref{table1}. The results vary by a factor of three.
 However, the highest value of the proton resonance width,  $2.96 \times 10^{-19}\,$ MeV,  was obtained using  Eq. (\ref{Ratioapprox2}), which is less accurate than the value of $1.87 \times 10^{-19}\,$  MeV obtained using Eq. (\ref{GammaANCmirrorratio3}).  However, this lower value of the proton resonance width is in an excellent agreement with the results obtained using the $R$-matrix approach.  
 
Note that we used two neutron ANCs. The higher one,  $C_{(n)\,0\,\frac{1}{2}} =   1.114$ fm${}^{-1/2}$, was obtained using the  SF obtained phenomenologically and from the shell-model calculations  \cite{Kolk, Okolowicz}, and the single-particle potential. The lower value of the ANC,   $C_{(n)\,0\,\frac{1}{2}} =   0.83$ fm${}^{-1/2}$, was obtained from the ${}^{10}{\rm B}(d,p){}^{11}{\rm B}(8.70\,{\rm MeV}; \frac{5}{2}^{+})$ reaction \cite{HindsMidleton,Fortune}. When using 
mirror symmetry method we assumed that the proton SF is equal to the neutron one.  The smallest  proton resonance width is  obtained  when we assume that the proton SF is equal to the neutron one deduced from  the deuteron stripping reaction. 

All the calculated proton resonance widths  are of the order of $10^{-19}$ MeV, in contrast to the phenomenological value $\Gamma_{(p)\,0\,\frac{1}{2}}= 2.3 \times 10^{-20}$ MeV deduced in \cite{Kolk}.  Note that a comprehensive fit was performed in \cite{Kolk} for higher energy data, $ 0.73 \leq E \leq 1.82\, {\rm MeV}$. Hence extracted in \cite{Kolk}, the proton resonance width of the near-threshold resonance  $E_{x}= 8.70$ MeV was found from the extrapolation of direct measurements to the low-energy region where direct data are unavailable. The difference between the proton resonance width calculated here and the phenomenological value deduced in \cite{Kolk} is distinct. 
The SMEC approach resulted in  $\,\Gamma_{(p)\,0\,\frac{1}{2}}= 2.2\times 10^{-17}$ MeV \cite{Okolowicz}, which is two orders of magnitude larger than our estimations and three orders of magnitude larger than the value obtained in \cite{Kolk}. 
  \hfill \break

\begin{widetext}
\begin{table}[htbp]
\caption{Proton  resonance width of the near-threshold resonance  $(\mathbf{8.70\,{\rm MeV};\frac{5}{2}^{+}})$   }
\begin{center}
\begin{tabular}{|c|c|c|}
\hline
Equations   & Mirror symmetry  & $R_{ch}\,$ (fm)     \\
\hline
                    &     $C_{(n)\,0\,\frac{1}{2}} =   1.114$ fm${}^{-1/2}$                    &                                               \\
\hline
&            $\Gamma_{(p)\,0\,\frac{1}{2} }\,$  (MeV)        &  \\
\hline
 (\ref{GammaANCmirrorratio3})  &    $1.87 \times 10^{-19}\,$  &       $\,4.0$ and $\,5.0\,$               \\
\hline 
(\ref{Ratioapprox2})   &    $2.96 \times 10^{-19}\,$     &      $\,4.6\,$       \\
\hline
                                   &     $C_{(n)\,0\,\frac{1}{2} } =   0.83$ fm${}^{-1/2}$    &   \\
  \hline   
      &            $\Gamma_{(p)\,0\,\frac{1}{2} }\,$  (MeV)        &  \\
\hline                          
 (\ref{GammaANCmirrorratio3})  &    $1.03 \times 10^{-19}\,$  &       $\,4.0$ and $\,5.0\,$       \\
\hline 
(\ref{Ratioapprox2})   &    $1.65 \times 10^{-19}\,$    &      $\,4.6\,$          \\
\hline    
&                        $R$-matrix approach  &            \\
\hline
   &            $\Gamma_{(p)\,0\,\frac{1}{2} }\,$  (MeV)        &  \\
\hline                          
(\ref{GpRmtrx1})               &                       $1.68 \times 10^{-19}\,$   &   $4.0$     \\   
 \hline                         
(\ref{GpRmtrx1})              &                      $1.86 \times 10^{-19}\,$  &    $5.0$         \\
 \hline   
 \label{table1}                                                              
\end{tabular}
\end{center}
\label{table1}
\end{table}
\end{widetext}

\section{\bf{Astrophysical  $S$-factor  for}  $\mathbf{{}^{10}{\rm B}}$- {\bf{proton fusion reaction}}}
 \label{Sfactor1}

In this section  the results of calculations of the $S$-factor for the low-energy ${}^{10}{\rm B}(p,\alpha){}^{7}{\rm Be}$ reaction are presented. We consider the energy region where the contribution from the near-threshold resonance at $E_{R}=0.01$ MeV dominates. Fitting to available low-energy data will allow us to determine the proton resonance width $\,\Gamma_{(p)\,0\,\frac{1}{2} }.$

We use the proton resonance widths $\,\Gamma_{(p)\,0\,\frac{1}{2}}$  determined from mirror symmetry and $R$-matrix method. Since we consider only very low energies, we took into account the contributions from the near-threshold resonance 
at $(.8.70\, {\rm MeV}; \frac{5}{2}^{+} )$  and the resonance $(9.20\, {\rm MeV}; \frac{5}{2}^{+})$, which interferes with the near-threshold resonance.  We also took into account the subthreshold bound state (aka subthreshold resonance)  $(8.65\,{\rm MeV}; \frac{7}{2}^{+})$  corresponding to the $(p\,{}^{10}{\rm B})$ binding energy $\varepsilon= 0.04$ MeV.
To extend our calculations up to $E= 0.27$ MeV following \cite{Spitaleri2018}  we included four negative parity resonances $(9.36\,{\rm MeV};\,\frac{5}{2}^{-}),\,$   $(9.645\,{\rm MeV}; \frac{3}{2}^{-}),\,$
$\,(9.83\,{\rm MeV}; \frac{5}{2}^{-})\,$ and  $(9.97\,{\rm MeV};\,\frac{7}{2}^{-}),\,$  which do not interfere with the near-threshold resonance. 
 
\subsection{\bf{Fitting to S-factor obtained from  direct measurements}}
\label{Angulofitting}
We begin with the $R$-matrix fitting to the $S$-factor from \cite{Angulo}. The main goal of our fit is to find the proton resonance width of the near-threshold resonance $(8.70\, {\rm MeV}; \frac{5}{2}^{+})$, which provides the best fit to the experimental data from \cite{Angulo} in the low-energy region where this resonance dominates.

When connsidering two interfering resonances,   $(8.70\, {\rm MeV}; \frac{5}{2}^{+})$ and $(9.20\,{\rm MeV}; \frac{5}{2}^{+}),\,$  we followed the procedure described in \cite{Barker}. The resonance energy of the former is taken to be equal to the first $R$-matrix energy eigenvalue and was used for the boundary condition energy, while the second  $R$-matrix energy level was varied to achieve the best fit. This procedure was explained in Appendix  \ref{Twointerferres1}. The second resonance determined from the search deviates from the experimental one. 
After that, using Barker's transformation \cite{Barker}, we can find the parameters of the second resonance.  

\begin{figure}[tbp]
\includegraphics*[scale=0.6]{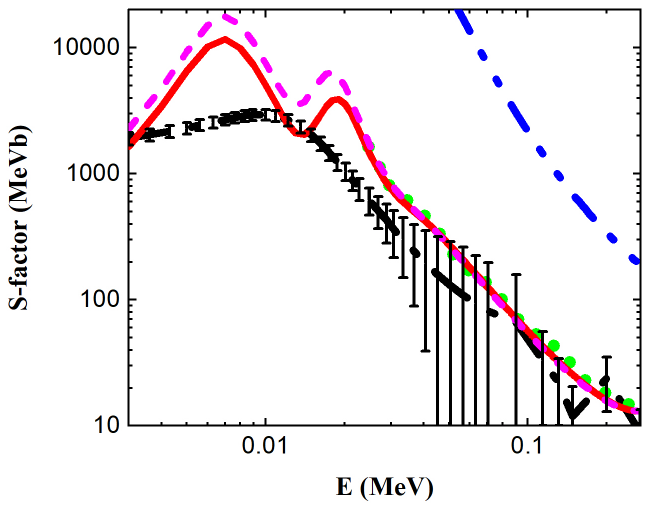}
\caption{Low-energy $S$-factors for the ${}^{10}{\rm B}(p,\alpha){}^{7}{\rm Be}$  reaction. 
The green short dots- $S$-factor from \cite{Angulo}. The black dash-dotted line with errors is the THM $S$-factor \cite{Spitaleri2018}. The solid red (magenta dash) line is the present work best fit to data from \cite{Angulo} for the proton resonance width  $\Gamma_{(p)\,0\,\frac{1}{2} }=1.68 \times 10^{-19}$ MeV 
( $\Gamma_{(p)\,0\,\frac{1}{2} }=2.5\times 10^{-19}$ MeV) of the resonance $E_{R}=0.01$ MeV. The blue  dash-dotted-dotted line is obtained with the proton resonance width $2.2\times 10^{-17}\,$MeV \cite{Okolowicz}.  }
\label{fig_Sfctrs}
\end{figure}

 Fig. \ref{fig_Sfctrs} depicts  the  low-energy $S$-factor for the  ${}^{10}{\rm B}(p,\alpha){}^{7}{\rm Be}$ reaction  from \cite{Angulo} compared with the present work  calculations. For comparison are also shown  the THM experimental $S$-factor and the $S$-factor calculated using the proton resonance $\,\Gamma_{(p)\,0\,1/2 }= 2.2 \times 10^{-17}\,$ Mev from \cite{Okolowicz}.  
The low-energy THM $S$-factor at $E < 0.034$ MeV becomes noticeably lower than the one from \cite{Angulo}. The $S$-factor calculated 
with the proton resonance width from \cite{Okolowicz}  reveals  irreconcilable difference  with other $S$-factors. 
 
 We find that the best fits to the $S$-factor from \cite{Angulo}  can be achieved using two different proton  resonance widths of the resonance $(8.70\, {\rm MeV}; \frac{5}{2}^{+})$  (see Table  \ref{table1}):  $\,\Gamma_{(p)\,0\,1/2}=1.68 \times 10^{-19}\,$ MeV   and
 $\,\Gamma_{(p)\,0\,\frac{1}{2}}=2.5 \times 10^{-19}\,$ MeV. These are the marginal values.
 Any resonance width between these two marginal values will provide a similar fit.

  The $R$-matrix parameters of our fits to the $S$-factor from \cite{Angulo}  are given in Table \ref{table_Rmatrix}.     Note that these parameters are taken from \cite{Spitaleri2018} with some modifications to achieve the best fit. In particular,  the proton and $\alpha$-particle resonance widths of the resonance $(E_{x}=9.36\,{\rm MeV}; \frac{5}{2}^{-})\,$ are varied in our fit.  Only six first resonances and one subthreshold resonance are taken into account. 
The channel radius in the proton channel  $R_{ch(p)} =4.0\,$fm and in the $\alpha$-channel  $R_{ch(\alpha)} =  4.49\,$fm are 
employed in these calculations.  
\begin{widetext}
 \begin{table}[htb]
\caption{The parameters of the $R$-matrix fit corresponding to the red solid line in Fig. \ref{fig_Sfctrs}}  
\begin{center}
\begin{tabular}{|c|c|c|c|c|c|c|}
\hline
$E_{x}\,{\rm MeV}$ &  $E_{R}\,$MeV  & $J^{\pi}$ & $l_{p}$ &    $l_{\alpha}$       &  $\Gamma_{(p)\,l_{p}\,j_{p} }\,$MeV & $\Gamma_{(\alpha)\,l_{\alpha}\,j_{\alpha}}\,$MeV  \\ 
\hline 
$8.654$ &  $-0.354$    &   $\frac{7}{2}^{+}$     &  $2$     &    $3$              &  $2.05 \times 10^{-27}$    &  0.005   \\
$8.70$ & $0.01$  & $\frac{5}{2}^{+}$    & $0$  &  $1$  &     $1.68 \times 10^{-19}$  \big[ $2.5 \times 10^{-19}$\big]          &     $0.015$\\
$9.36$ & $0.671$  & $\frac{5}{2}^{-}$    & $1$  &  $2$  &     $ 0.085$   \big[$0.090$\big]          &     $0.350$                                  \\
$9.645$    & $0.956$      &  $\frac{3}{2}^{-}$       &   $1$     &    $0$            &       $0.048$            &      $0.223$     \\
 $9.83$     & $1.140$      &  $\frac{5}{2}^{-}$       & $1$      &     $2$            &       $0.012$         &         $0.165$      \\
 $9.97$     &   $1.284$      &  $\frac{7}{2}^{-}$       & $1$      &   $2$           &      $0.221$          &         $0.066$      \\
 $12.98$ \big[13.693 \big] & $4.29$ \big[$5.00$\big] & $\frac{5}{2}^{+}$    & $0$  &  $1$  &     $0.114$  \big[$0.1303$\big]           &     
 $ 0.018$   \big[$0.015$\big]   \\
\hline   
\end{tabular}
\end{center}
\label{table_Rmatrix}
\end{table} 
\end{widetext}
In Table \ref{table_Rmatrix} all the numbers in the brackets  correspond to the resonance width $\Gamma_{(p)\,0\,\frac{1}{2} }=2.5\times 10^{-19}$ MeV.  
As explained above,  in the $R$-matrix fit with two interfering resonances and fixed one of them, namely, the resonance $ (8.70\,{\rm MeV}; \,\frac{5}{2}^{+}),\,$ the second resonance   $(9.20\,{\rm MeV}; \,\frac{5}{2}^{+})\,$ is replaced with the resonance which is varied. The obtained eigenenergy for the second level  is  $E_{2}= -0.115$ MeV, which is the same for both widths $\Gamma_{(p)\,0\,\frac{1}{2} }$ listed in Table \ref{table_Rmatrix}.
 After applying the Barker's procedure \cite{Barker}  the second resonances are  $4.290$ MeV for $\,\Gamma_{(p)\,0\, \frac{1}{2}}=1.68 \times 10^{-19}\,$ MeV and $5.00$ MeV for  $\,\Gamma_{(p)\,0\, \frac{1}{2}}=2.5 \times 10^{-19}\,$ MeV. They are given in the last row of Table \ref{table_Rmatrix}. These resonances significantly deviate from the experimental resonance at $0.51$ MeV. It means that  
there is a significant contribution from the  $\frac{5}{2}^{+}$  background resonance, which was not explicitly included at the beginning but revealed itself through the fit. Since initially we took into account only two $\frac{5}{2}^{+}$ resonances, the effective second resonance plays the role of the background one with the proton and alpha-particle partial resonance widths given in Table \ref{table_Rmatrix}. More sophisticated fit should take into account three interfering $\frac{5}{2}^{+}$ resonances: $8.70, \,$  $9.20$ MeV and the background resonance. 
An important contribution at higher energies (in our fitted energy interval) gives the resonance $\,(9.36\,{\rm MeV}; \,\frac{5}{2}^{-})\,$ discovered in \cite{Spitaleri2018}, which was not listed in compilation \cite{Kelley}.
 Our $S$-factors at $E_{R}=0.01\,$MeV correspond to the cross sections: $\sigma(0.01) = 1.38 \times 10^{-12}$ mb 
 for $\,\Gamma_{(p)\,0\,\frac{1}{2}}=1.68 \times 10^{-19}\,$ MeV  and  $\sigma(0.01) = 2.05 \times 10^{-12}$ mb 
 for  $\,\Gamma_{(p)\,0\,\frac{1}{2}}=2.5 \times 10^{-19}\,$ MeV.

 The lowest measured energy $0.0246$ MeV in \cite{Angulo}  is higher than the near-threshold resonance  $0.01$ MeV. It is also the NIF operational energy. However, from Fig. \ref{fig_Sfctrs}  follows that the $S$-factor is sensitive to the variation of the proton resonance width $\,\Gamma_{(p)\,0\,\frac{1}{2}}$ only at energies $E<0.021$ MeV. That is why the analysis of the data from \cite{Angulo}  allowed us to identify a broad range of the proton resonance widths rather than a unique value. 

Only the indirect THM reached energies close to $0.01$ MeV. 
In the next subsection, we present the results of the fitting to the THM  $S$-factor \cite{Spitaleri2018}.

 \subsection{\bf{Fitting THM data}  }
 \label{THMfit}
 
\begin{figure}[tbp]
\includegraphics*[scale=0.57]{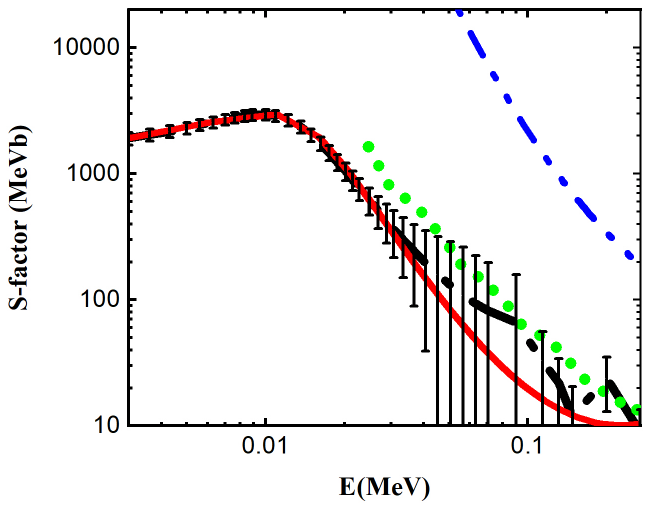}
\caption{The low-energy $S$-factors for the ${}^{10}{\rm B}(p,\alpha){}^{7}{\rm Be}$  reaction. 
All the notations  are the same as in Fig. \ref{fig_Sfctrs}  except  for  the solid red line, which is now  our $R$-matrix fit to the THM  
data from \cite{Spitaleri2018}. }
\label{fig_THMMSfctr1}
\end{figure}

  There were a few publications in which THM  measurements of the low-energy $S$-factor for the ${}^{10}{\rm B}(p,\alpha){}^{7}{\rm Be}$  fusion were reported  \cite{Spitaleri2014,Spitaleri2017,Spitaleri2018}. In this fit, we use the data from \cite{Spitaleri2018}.
 The fit is similar to the one described in subsection \ref{Angulofitting}.  
 The result of the fit is shown in Fig. \ref{fig_THMMSfctr1}.
Since the THM  $S$-factor is lower than the  one from \cite{Angulo}, we are able to determine  the proton resonance width  $\Gamma_{(p)\,0\,\frac{1}{2}}=1.0 \times 10^{-19}\,$ 
MeV providing the best fit. This width is smaller than the proton resonance width interval deduced from the fit of the $S$-factor to data from \cite{Angulo}. Our $S$-factor perfectly fits the experimental THM one.
In particular,  our fit confirms the bending of the $S$-factor at energies below the near-threshold resonance $0.01$ MeV. The behavior of the THM $S$-factor significantly deviates from the fit to data from \cite{Angulo} extrapolated down to energies $E < 0.021$ MeV.
 
  The $R$-matrix amplitude  has two $\frac{5}{2}^{+}$  interfering 
 resonances,  $8.70$ and $\,9.20\,$ MeV.  Again, as we did for fitting the $S$-factor from \cite{Angulo},  we fix the former 
 and vary the energy of the second resonance. 
 The $R$-matrix parameters of the fit to the THM $S$-factor from \cite{Spitaleri2017}  are given in  Table \ref{table_RmatrixTHM}.

  \begin{table}[htb]
\caption{The parameters of the $R$-matrix fit shown as the red solid line in Fig. \ref{fig_THMMSfctr1} } 
\begin{center}
\begin{tabular}{|c|c|c|c|c|c|c|}
\hline
$E_{x}\,{\rm MeV}$ &  $E_{R}\,$MeV  & $J^{\pi}$ & $l_{p}$ &    $l_{\alpha}$       &  $\Gamma_{(p)\,l_{p}\,j_{p} }\,$MeV & $\Gamma_{(\alpha)\,l_{\alpha}\,j_{\alpha}}\,$MeV  \\ 
\hline 
$8.654$ &  $-0.354$    &   $\frac{7}{2}^{+}$     &  $2$     &    $3$              &  $2.05 \times 10^{-27}$    &  0.005   \\
$8.70$ & $0.01$  & $\frac{5}{2}^{+}$    & $0$  &  $1$  &     $1.00\, \times 10^{-19}$             &     $0.015$\\
$9.36$ & $0.671$  & $\frac{5}{2}^{-}$    & $1$  &  $2$  &     $ 0.05$             &     $0.450$                                  \\
$9.645$    & $0.956$      &  $\frac{3}{2}^{-}$       &   $1$     &    $0$            &       $0.048$            &      $0.230$     \\
 $9.83$     & $1.140$      &  $\frac{5}{2}^{-}$       & $1$      &     $2$            &       $0.012$         &         $0.165$      \\
 $9.86$ & $1.17$  & $\frac{5}{2}^{+}$    & $0$  &  $1$  &     $ 0.004$     &     $ 0.239$\\
 $9.97$     &   $1.284$      &  $\frac{7}{2}^{-}$       & $1$      &   $2$           &      $0.221$          &         $0.066$      \\
\hline   
\end{tabular}
\end{center}
\label{table_RmatrixTHM}
\end{table}
The channel radii are $R_{ch(p)}=4.0\,$fm and $R_{ch(\alpha)}=  4.05\,$ fm. As the result of the fit, the second resonance
$9.20$ MeV is replaced with the resonance level $9.86$ MeV. This resonance takes into account both $9.20$ MeV resonance and 
the background one.  It is worth mentioning that the resonance  at $9.36$ MeV is important for  fitting at energies close to $0.27$ MeV.

Our $S$-factor at $E_{R}=0.01\,$MeV corresponds to the cross-section for the ${}^{10}{\rm B}(p,\alpha){}^{7}{\rm Be}$  reaction: $\sigma(0.01) = 1.38 \times 10^{-12}$ mb 
 for $\,\Gamma_{(p)\,0\,\frac{1}{2}}=1.0 \times 10^{-19}\,$ MeV.

\section{\bf{Conclusion}}
The extra low-energy ${}^{10}{\rm B}(p,\alpha){}^{7}{\rm Be}$   reaction  is dominated  by the near-threshold resonance at $(8.70\,{\rm MeV};\,\frac{5}{2}^{+})$. The cross-section of this reaction is defined by the proton resonance width $\,\Gamma_{(p)\,0\,\frac{1}{2} }\,$  in the entry channel and by the $\alpha$- channel resonance width in the exit channel. The latter is well established, while the former is unknown due to its smallness. 
This paper presents two methods of estimating the proton resonance width: mirror symmetry and $R$-matrix. 
The first one gave four different values in the interval $(1.03 -2.5) \times 10^{-19}\,$ MeV, while the application of the second method resulted in   $\Gamma_{(p)\,0\,\frac{1}{2}} = 1.68 \times 10^{-19}\,$ and  $1.86 \times 10^{-19}\,$MeV.   These widths are almost an order of magnitude larger than the phenomenological width deduced in \cite{Kolk}, but two orders of magnitude smaller than the width from \cite{Okolowicz}.
 
 Using the found proton widths, we achieved excellent agreements with experimental low-energy $S$ factors from \cite{Angulo} and \cite{Spitaleri2018} in the energy interval $0.001 < E < 2.7\,$ MeV. An important contribution to the $S$-factor at energies near $\,2.7\,$ MeV gives the negative parity resonance at $\,(9.36\,{\rm MeV}; \,\frac{5}{2}^{-})\,$ MeV found in  \cite{Spitaleri2018}. 
Since the lowest measured energy in \cite{Angulo} was $0.0246$, which is higher than the near-threshold resonance $0.01$ MeV,  we were able to identify the interval of the resonance widths providing the best fit rather than a unique value: $\,\Gamma_{(p)\,0\,\frac{1}{2} }= (1.68 - 2.5) \times 10^{-19}$ MeV. The THM experiment  \cite{Spitaleri2018} reached  $0.01$ Me and we were able to determine a unique value of the proton resonance width:  $\,\Gamma_{(p)\,0\,\frac{1}{2} }= 1.0 \times 10^{-19}$ MeV.

From the fits we find the cross-sections for the ${}^{10}{\rm B}(p,\alpha){}^{7}{\rm Be}$   reaction at $E=0.01$ MeV:  $\sigma(0.01) = 1.38 \times 10^{-12}$ mb  for $\,\Gamma_{(p)\,0\,\frac{1}{2}}=1.68 \times 10^{-19}\,$ MeV,  $\sigma(0.01) = 2.05 \times 10^{-12}$ mb 
 for  $\,\Gamma_{(p)\,0\,\frac{1}{2}}=2.5 \times 10^{-19}\,$ MeV (fits to the $S$-factor \cite{Angulo})  and   $\sigma(0.01) = 1.38 \times 10^{-12}$ mb  
 for $\,\Gamma_{(p)\,0\,\frac{1}{2}}=1.0 \times 10^{-19}\,$ MeV  (fit to the THM $S$-factor \cite{Spitaleri2018}). These values can help to determine whether 
 the  ${}^{10}{\rm B}(p,\alpha){}^{7}{\rm Be}$ reaction is a spoiler for the  
 ${}^{11}{\rm B}(p,\,2\,\alpha){}^{4}{\rm He}$ reaction used by NIF for the energy production.

 The higher-energy $S$ factor was not calculated because of the uncertainties in the location and spin-parity assignments of low-energy resonances \cite{Kolk,Spitaleri2018}.   It calls for further investigation of low-energy resonances
 in ${}^{11}{\rm C}$.

\acknowledgments{The author acknowledges that this material is based upon work supported by the US DOE National Nuclear Security Administration, under Award Number DE-NA0003841 and  DOE Grant No. DE-FG02-93ER40773. The author  thanks Michael Wiescher and Richard de Boer for useful  discussions. The author expresses his gratitude to Arkadi Zhanov for technical support. }

\section{\bf{Appendix}}

\numberwithin{equation}{section}
\begin{appendix}

\section{\bf{Internal resonance wave function in the $R$-matrix approach}}
\label{intreswf1}

We start from Eq. (107) \cite{mukRes2023} generalizing it for resonance states. Below we use  simplified notations leaving only $l$-dependence of the wave functions.
 The resonance state can be  Gamow-Siegert
resonance wave function. For narrow resonance the imaginary part of the resonance energy can be neglected  and the internal Gamow-Siegert wave function becomes a real, regular solution in the nuclear interior with outgoing wave $O_{l}$. 
Introducing Zel'dovich regulator  $\lim\limits_{\beta \to 0} e^{-\beta\,r^{2}}$   one can regularize  the Gamow-Siegert  wave function. Hence we can assume that the integral 
\begin{align}
&J_{0}^{\infty}=\int\limits_{0}^{\infty}\,{\rm d}r \phi_{l}^{2}(k,r)
\label{normint1}
\end{align}
is finite.
 
Then  Eq. (107) \cite{mukRes2023}   becomes valid  for  the resonance state 
\begin{align}
&J_{0}^{\infty}=\int\limits_{0}^{\infty}\,{\rm d}r \phi_{l}^{2}(k,r)=
 \Bigg[ 1 + {\int\limits_{R_{ch}}^{\infty}\,{\rm d}r \varphi_{l}^{2}(k,r)}\Bigg]\,\int\limits_{0}^{R_{ch}}\,{\rm d}r \phi_{l}^{2}(k,r)   \nonumber\\
&=\Big[ 1+  ({\tilde \gamma}_{l}^{sp})^{2}\, \frac{\partial  {\hat S}(E)}{\partial E}\Big]\,\int\limits_{0}^{R_{ch}}\,{\rm d}r \phi_{l}^{2}(k,r).
\label{NR1}
\end{align}
Here 
\begin{align}
\varphi_{l}(k,r)  =  \frac{\phi_{l}(k,r)}{\Big [\int\limits_{0}^{R_{ch}}\,{\rm d}r\,\phi_{l}^{2}(k,r)\Big]^{1/2}}
\label{tildeul1}
\end{align}
is  the resonant wave function normalized to unity over the internal region ($r \leq R_{ch}$),  that is, 
\begin{align}
\int\limits_{0}^{R_{ch}}\,{\rm d}r\,\varphi_{l}^{2}(k,r)= 1.
\label{inernnorm1}
\end{align}
The boundary condition is
\begin{align}
&R_{ch}\,\frac{ {\rm d}\ln \big(\phi_{l }(k,r)\big)}{{\rm d}r}\Big|_{r=R_{ch}} =  R_{ch}\,\frac{ {\rm d}\ln \big(O_{l }(k,r)\big)}{{\rm d}r}\Big|_{r=R_{ch}}
\nonumber\\
&= {\hat S}(E),
\label{logderbs1}
\end{align}
where ${\hat S}(E)$ is the energy shift function.
The condition (\ref{logderbs1})  in which the wave function in the external region is giving by the outgoing wave assumes that $k=k_{0}$, where
$k_{0}$ is the real  part of the resonance momentum related to the real part of the  resonance energy as $E_{R}= k_{0}^{2}/(2\mu)$.   Hence in what follows to underscore explicitly that
we deal with the resonance state,  we replace $k$ with $k_{0}$. 
Assume that $\phi_{l }(k_{0},r)$ is normalized to unity over the entire configuration space. Then 
\begin{align}
&\phi_{l}^{2}(k_{0},r)  = 
\frac{\varphi_{l}^{2}(k_{0},r) }{1 + ({\tilde \gamma}_{l}^{sp})^{2}\,\frac{\partial  {\hat S}}{\partial E}\Big|_{ E= E_{R}}}      \nonumber\\
&= 2\,\mu\,R_{ch}\,\frac{ ({\tilde \gamma}_{l}^{sp})^{2} }{1 + ({\tilde \gamma}_{l}^{sp})^{2}\,\frac{\partial  {\hat S}}{\partial E}\Big|_{ E= E_{R}}}
\label{phyu1}
\end{align}
with
\begin{align}
\varphi_{l}^{2}(k_{0},R_{ch}) = 2\,\mu\,R_{ch}\,({\tilde \gamma}_{l}^{sp})^{2}.
\label{varphiintrw1}
\end{align}
$({\tilde \gamma}_{l}^{sp})^{2}$  is the formal $R$-matrix single-particle reduced  width and  
\begin{align}
(\gamma_{l}^{sp})^{2}  =     \frac{ ({\tilde \gamma}_{l}^{sp})^{2} }{1 + ({\tilde \gamma}_{l}^{sp})^{2}\,\frac{\partial  {\hat S}}{\partial E}\Big|_{ E= E_{R}}}
\label{obsredwampl1}
\end{align}
is the observable single-particle reduced width. Then 
\begin{align}
&\phi_{l}^{2}(k_{0},R_{ch})=  2\,\mu\,R_{ch}\,(\gamma_{l}^{sp})^{2}.
\label{varphiobsspamma1}
\end{align}
Also 
\begin{align}
\gamma_{l}^{2} =   {\mathtt  S}_{l}\, (\gamma_{l}^{sp})^{2}
\label{obsredw1}
\end{align}
is the observable reduced width and  $\,{\mathtt  S}_{l}\,$  is the SF in the single-particle approximation.

\section{\bf{Two interfering resonances in $R$-matrix approach}}
\label{Twointerferres1}

Here we consider the application of the $R$-matrix  method for two interfering resonances  following \cite{Barker,Barker2008}.
The expression of the $S$-factor for two-level, two-channel case is given by
\begin{align}
& S(E)(MeVb)= \frac{{\hat J}_{R}}{{\hat J}_{a}\,{\hat J}_{A}}\,\lambda_{N}^{2}\,\,m_{au}^{2}\,e^{2\,\pi\,\eta_{i}}\,\frac{2\,\pi}{\mu_{i}} \,10^{-2}\,    \nonumber\\                                                                                
& \times \,P_{f}(E_{f})\,P_{i}(E)\,\Big|\sum\limits_{\nu\,\tau}\,{\tilde \gamma}_{f\,\nu}\,\big[{\rm {\bf {\cal A}}}^{-1}\big]_{\nu\,\tau}(E)\,{\tilde \gamma}_{i\,\tau}\,\Big|^{2},
\label{Sfactr2l2ch1}
\end{align}
where $J_{R}$ is the spin of the resonance, $J_{i}$ is the spin of particle $i$, ${\hat J}_{i}=2\,J_{i}+1$,  $\lambda_{N}$ is the Compton
wavelength, $m_{au}$ is the atomic mass unit, $\mu_{i}$ and $\eta_{i}$ are the reduced mass and the Coulomb parameter of the particles $a$ and $A$ in the initial channel,
$P_{i}$ and $P_{f}$ are the penetrability factors in the channels $i$ and $f$, ${\tilde \gamma}_{c\,\nu}$ is the formal reduced width amplitude in the channel $c=i,\,f$ and level $\nu=1,2$. $\,E_{c}$ is the relative kinetic energy of the particles in channel $c$ and $E \equiv E_{i}$. 
We assume that the SFs  for levels $\nu=1,2$ are equal to unity.
 The  level matrix in the  $R$-matrix method is defined as
\begin{align}
& \big[{\rm {\bf {\cal A}}}\big]_{\nu\,\tau}(E)= (E_{\nu} - E)\,\delta_{\nu\,\tau} -  \nonumber\\
\sum\limits_{c=i,f}\,{\tilde \gamma}_{c\,\nu}\,
&{\tilde \gamma}_{c\,\tau}\,\big[{\hat S}_{c}(E_{c}) - B_{c} + i\,P_{c} \big]. 
\label{ARmatr1}
\end{align}
Here  ${\hat S}_{c}(E_{c})\,$ is the energy shift in the channel $c=i,\,f$ (see Eq. (\ref{logderbs1})),  and  $\,B_{c}\,$  is the level-independent boundary condition in the channel $c$ and $E_{\nu}$ is the eigenenergy of level $\nu=1,2$, which is the $R$-matrix fitting parameter.  

In the fit it is convenient to take one of the level eigenenergies equal to the resonance energy of the corresponding level, 
for example $E_{1} = E_{R_{1}}$ and $B_{c}={\hat S}_{c}(E_{R_{1}}).\,$   $E_{R_{1}}$ is the real part of the resonance energy of the level $1$. Then the second eigenvalue $E_{2}$ is the fitting parameter, which deviates from the resonance energy $E_{R_{2}}$ of the second level.  After that, using  the Barker's transformation \cite{Barker} one can find the energy of the second resonance.

\end{appendix}

\end{document}